\documentclass[letterpaper]{article} 
\usepackage{aaai23}  
\usepackage{times}  
\usepackage{helvet}  
\usepackage{courier}  
\usepackage[hyphens]{url}  
\usepackage{graphicx} 
\usepackage{natbib}  
\usepackage{caption} 
\usepackage{algorithm}
\usepackage{algorithmic}

\usepackage{newfloat}
\usepackage{listings}

\usepackage{amssymb}
\usepackage{booktabs}
\usepackage{makecell}
\usepackage{subfigure}

\DeclareCaptionStyle{ruled}{labelfont=normalfont,labelsep=colon,strut=off} 
\floatstyle{ruled}
\newfloat{listing}{tb}{lst}{}
\floatname{listing}{Listing}
\title{Layer Imbalance Aware Multiplex Network Embedding}
\author {
	Kejia Chen,\textsuperscript{\rm 1}
	Yinchu Qiu, \textsuperscript{\rm 2}
	Zheng Liu \textsuperscript{\rm 3}
}
\affiliations {
	\textsuperscript{\rm 1} chenkj@njupt.edu.cn\\
	\textsuperscript{\rm 2} 1221045505@njupt.edu.cn\\
	\textsuperscript{\rm 3} zliu@njupt.edu.cn\\
}

\usepackage{bibentry}

\begin{document}

\maketitle

\begin{abstract}
Multiplex network embedding is an effective technique to jointly learn the low-dimensional representations of nodes across network layers. However, the number of edges among layers may vary significantly. This data imbalance will lead to performance degradation especially on the sparse layer due to learning bias and the adverse effects of irrelevant or conflicting data in other layers. In this paper, a Layer Imbalance Aware Multiplex Network Embedding (LIAMNE) method is proposed where the edges in auxiliary layers are under-sampled based on the node similarity in the embedding space of the target layer to achieve balanced edge distribution and to minimize noisy relations that are less relevant to the target layer. Real-world datasets with different degrees of layer imbalance are used for experimentation. The results demonstrate that LIAMNE significantly outperforms several state-of-the-art multiplex network embedding methods in link prediction on the target layer. Meantime, the comprehensive representation of the entire multiplex network is not compromised by the sampling method as evaluated by its performance on the node classification task.
\end{abstract}

\section{Introduction}
Networks or graphs are often used to visually represent relations between objects in complex systems. Each node in a given network can be embedded as a low-dimensional vector through analysis and learning techniques, to facilitate downstream tasks of classification and inference. The prevailing network embedding learning methods include DeepWalk \cite{c:1}, node2vec \cite{c:2}, GNNs \cite{c:3}, etc., which mainly deal with single-layer networks with a single relation type between nodes.

Relations in real networks are more complex and diverse, usually formalized as multiplex networks. For example, interactions between users on Twitter may have multi-types of relations such as \emph{like}, \emph{reply} and \emph{retweet}. Each type of relational data is constructed as a single-layer network, where nodes are users and edges correspond to a type of user-to-user behaviors. According to the number of node embedding spaces, multiplex network embedding (MNE) methods can be classified into two categories \cite{c:4}: Joint Representation Learning (JRL) and Coordinated Representation Learning (CRL). 
JRL methods aim to combine node representations from multiple layers into one shared feature space, while CRL methods intend to learn separated node embeddings so that each layer has an independent feature space. This paper adopts the methodology of CRL to better learn the semantics information for each layer, especially for the sparse target layer.

However, there often exists serious data imbalance among multiple layers. Taking Twitter as an example, the number of \emph{retweets} is far less than that of \emph{reads} or \emph{follows}. Directly using existing MNE methods on layer-imbalanced multiplex networks may lead to the following two problems: 1) the node representation in sparse layers may not be well learned. For example, models are likely to over-learn browsing behavior during training while under-learning retweeting behavior, although the latter better reflects preferences among users. 2) some edges in other layers that are less correlated with the target layer may be distractions for learning node representation on the target layer. For example, browsing a user's tweet does not necessarily result in a retweet.

To solve the above problems, this paper proposes a novel MNE method called Layer Imbalance Aware Multiplex Network Embedding (LIAMNE). The method is a CRL method that learns a common embedding and multiple layer embeddings for each node. Firstly, the base layer embeddings on the original multiplex network are obtained using a baseline embedding method. Secondly, the edges in auxiliary layers are under-sampled based on node similarities from the target layer, which not only achieves a balance among layers but also removes noisy relations that are less relevant to the target layer. Finally, coordinated representation learning is performed on a relatively balanced multiplex network after sampling. 

The main contributions of LIAMNE are as follows:

\begin{itemize}
    \item To the best of our knowledge, we make the first attempt to solve the layer imbalance problem in multiplex network embedding.
    \item We propose a novel under-sampling method on auxiliary layers to improve the quality of node embeddings on target layer.
    \item Experimental results show that our model significantly outperforms several benchmark MNE models in both link prediction on the target layer and node classification.
\end{itemize}

\section{Related Work}
\subsection{Multiplex Network Embedding}
Multiplex network embedding (MNE) can be implemented via Joint Representation Learning (JRL) or Coordinated Representation Learning (CRL). JRL is to combine node representations from multiple layers into one embedding which means all node embeddings share the same feature space. DPMNE \cite{c:6} is a typical JRL method that learns node embeddings by simultaneously minimizing the deep reconstruction loss with the autoencoder neural network, enforcing the data consistency across views via common latent subspace learning.

Unlike JRL, CRL learns an independent feature space for each layer. It becomes the dominant MNE methodology as it can combine the information from different types of relations while maintaining their distinctive properties. MNE \cite{c:7} and PMNE \cite{c:8} propose to learn one high-dimensional common embedding and a lower-dimensional additional embedding for each type of relation. GATNE \cite{c:5} borrows ideas from both JRL and CRL to learn a base embedding and multiple edge embeddings (i.e. layer embedding in this paper) for each node separately. The final embedding of GATNE is a combination of both embeddings with a self-attention mechanism. CrossMNA \cite{c:9} leverages the cross-network information to refine two types of node embeddings: inter-vector for network alignment and intra-vector for other downstream network analysis tasks. 
HDMI \cite{c:22} and DGMI \cite{c:21} extend DGI \cite{c:20} on multiplex networks. Both are CRL methods since they learn node embeddings for each layer.

To better learn the semantics information on each layer, especially the sparse target layer, this paper adopts the CRL method.

\subsection{Network Imbalance}
Currently, the network imbalance problem mainly refers to the imbalance of node label distribution and the proposed solutions include GraphSMOTE \cite{c:10}, NRGNN \cite{c:11}, FRAUDRE \cite{c:12}, etc. 

The imbalance between relation types was once discussed in heterogeneous network embedding. BHIN2vec \cite{c:14} proposed a random-walk strategy that generates training samples according to the relative training ratio, which results in a balanced training for the node embedding. Similarly, PME \cite{c:15} proposed a novel loss-aware adaptive sampling approach for model optimization. However, the above methods for heterogeneous networks fail to capture the cross-relational information, i.e., cross-layer information in multiplex networks. In the research community of MNE, only CrossMNA \cite{c:9} mentioned the problem of data imbalance among multiple layers but it mainly focuses on the imbalance of anchor links in multi-networks alignment. To our best knowledge, the layer imbalance (i.e., the imbalance of relation types) problem in MNE has not received extensive attention. 

In this paper, an under-sampling method based on node similarity is proposed to obtain a layer-balanced multiplex network thereby enhancing the node embeddings on the sparse target layer.

\section{PROBLEM DEFINITION}
DEFINITION 1 (Multiplex Networks). A multiplex network is a network $G=(V, E), E= \cup_{l\in L}E_l(|L|>1)$,
where $E_l$ consists of all edges on the $l$th layer of the network, that is $G_l=(V, E_l)$. 
~\\
~\\
DEFINITION 2 (Layer Imbalance). A ratio $\mu ={\rm log} \frac{|E_{max}|}{|E_{min}|}$ is defined to roughly measure the layer imbalance of a multiplex network $G$, where $|E_{max}|$, $|E_{min}|$ represent the number of edges on the densest layer $G_{max}$ and on the sparsest layer $G_{min}$, respectively. The layers of $G$ are more imbalanced when the ratio $\mu$ is larger. 
~\\
~\\
PROBLEM 1 (Multiplex Network Embedding). Given a multiplex network $G=(V,E)$ where the $l$th layer $G_l=(V,E_l)$. The problem of multiplex network embedding is to learn low-dimensional representations for each node in $V$, which can be implemented via two ways, i.e., JRL and CRL. The JRL methods learn a mapping function $f_{\theta}:V\rightarrow \mathbb{R}^{|V|\times d}$ to embed the nodes of $G$ into one feature space, while the CRL methods try to find a map function $f_{\theta_l}:V\rightarrow \mathbb{R}^{|V|\times d}$ for each layer $G_l$.
~\\
~\\
PROBLEM 2 (Link Prediction in Multiplex Networks). Given two nodes $v_i$ and $v_j$ on $G_l$. The problem of link prediction in multiplex networks usually refers to inferencing whether there is an edge between $v_i$ and $v_j$ on a given layer $G_l$. In this paper, the layer that performs link prediction is called the target layer $G_t$, which is a sparse network in most cases, and the other layers are called the auxiliary layer $G_a$.

\begin{figure*}[t]
\centering
\includegraphics[width=500pt]{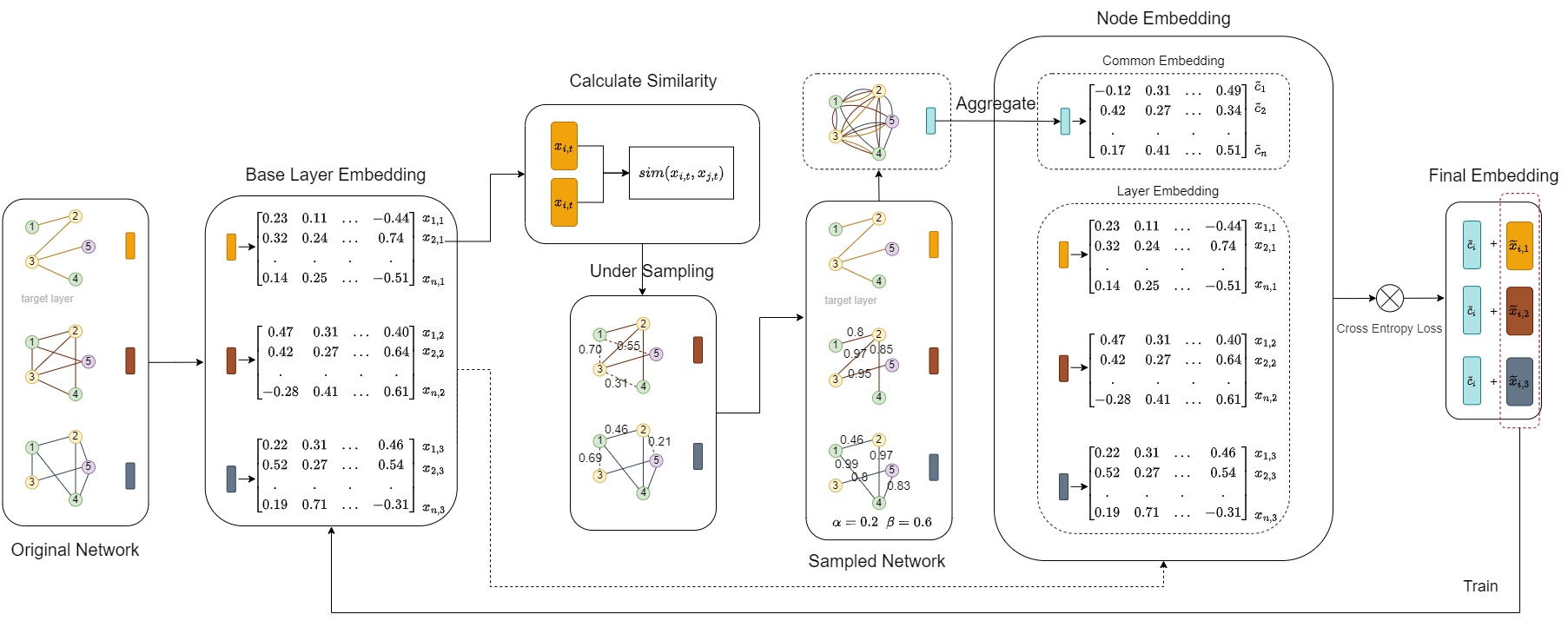}
\caption{Overview of LIAMNE. For each node, two types of embeddings are finally obtained by the training phrase: 
common embedding $\widetilde{c}$ for entire networks and layer embedding $\widetilde{x}_l$ for each layer, 
and the embedding $\widetilde{c}+\widetilde{x}_l$ can be used for downstream tasks.}\label{fig:1}
\end{figure*}

\section{LIAMNE}
In this section, we propose a CRL-based method named Layer Imbalance Aware Multiplex Network Embedding (LIAMNE), which contains three modules: base layer embedding module, under-sampling module and final embedding module. The overall framework is shown in Figure \ref{fig:1} and each module will be detailed as follows.

\subsection{Base Layer Embedding}
The first step of LIAMNE is to learn the base layer embeddings using a baseline node embedding method, which is the preparation for under-sampling in auxiliary layers. For node $v_i$, its layer embedding $x_{i,l}$ from layer $G_l$ can be obtained by applying a random walk-based method, such as node2vec \cite{c:2}, or by aggregating its neighbors on $G_l$ through mean aggregation or other pooling aggregation operations. 

In this paper, a straightforward way is utilized. The base layer embeddings in the $k^{th}$th epoch are initialized by the updated layer embeddings after gradient descent in the ${(k-1)}^{th}$ epoch. Preliminary experiments in this paper show that this simple method achieves similar results but faster training speed compared with complex GNN methods. More details of the base layer embedding implementation can be seen in the Experimental Configuration section. 

\subsection{Under-Sampling on Auxiliary Layers}
The purpose of this module is to under-sampling auxiliary layers to retain as much as possible the complementary information that is beneficial to the target layer and finally generate a relatively balanced multiplex network. 

Given an auxiliary layer to be sampled $G_m=(V, E_m)$ and a target layer $G_t=(V, E_t)$, a sampling function $f_{s}(G_m, G_t)\rightarrow \widetilde{G}_m$ is defined to obtain a new layer $\widetilde{G}_m$.
First of all, a similarity function of nodes on the target layer is calculated and defined as the sampling probability of all edges on auxiliary layers:

\begin{equation}
\label{eq:1}
    {\rm sim}(v_i, v_j) = \frac{1}{1 + e^{-x_{i,t} x_{j,t}}}, j\neq i,
\end{equation}
where $x_{i,t}$ and $x_{j,t}$ represent the embeddings of $v_i$  and $v_j$ on the target layer, respectively. 

Then, two thresholds $\alpha$ and $\beta$ are set during sampling. 
Specifically, the edge $(v_i, v_j)_m$ will be removed directly if ${\rm sim}(v_i, v_j)<\alpha$ 
and be sampled if ${\rm sim}(v_i, v_j)>\beta$. If ${\rm sim}(v_i, v_j)\in [\alpha, \beta]$, 
it will be directly used as the sampling probability $p_s$, i.e. $p_s={\rm sim}(v_i, v_j)$ to decide whether sample $(v_i,v_j)_m$ or not. 

The pseudo code of the under-sampling process is listed in Algorithm \ref{alg:1}.
\begin{algorithm}[tb]
\caption{Under-sampling on auxiliary layers}
\label{alg:1}
\textbf{Input}: Auxiliary layer to be sampled $G_m=(V, E_m)$, 
target layer $G_t=(V, E_t)$, similarity threshold $\alpha$, $\beta$.\\
\textbf{Output}: Sampled auxiliary layer $\widetilde{G}_m=(V, \widetilde{E}_m)$.\\
\begin{algorithmic}[1] 
\STATE Let $\widetilde{E}_m=\emptyset$.
\WHILE{edge $(v_i, v_j)$ in $E_m$}
\IF {$(v_i, v_j)\in E_t$}
\STATE $\widetilde{E}_m\leftarrow \widetilde{E}_m\cup (v_i, v_j)$.
\ELSE
\IF {${\rm sim}(v_i, v_j)\in [\alpha, \beta]$}
\STATE $\widetilde{E}_m\leftarrow \widetilde{E}_m\cup (v_i, v_j)$ at a probability $p_s$. 
\ELSIF {${\rm sim}(v_i, v_j)>\beta$}
\STATE $\widetilde{E}_m\leftarrow \widetilde{E}_m\cup (v_i, v_j)$.
\ENDIF
\ENDIF
\ENDWHILE
\STATE \textbf{return} $\widetilde{G}_m=(V, \widetilde{E}_m)$
\end{algorithmic}
\end{algorithm}

\subsection{Final Embedding}
After under-sampling, the multiplex network $G=(V,E)$ is transformed into $\widetilde{G}=(V, \widetilde{E})$, where the final embedding $\widetilde{z}_{i,l}$ for node $v_i$ on the layer $\widetilde{G}_l$ is obtained which consists of two parts: common embedding $\widetilde{c}_i$ and layer embedding ${\widetilde{x}_{i,l}}$. The common embedding $\widetilde{c}_i$ represents the global information of $v_i$ shared among multiple layers. The layer embedding $\widetilde{x}_{i,l}$ preserves the structural features of $v_i$ in the layer network ${\widetilde{G}_l}$ which cannot be shared with other layers. 

To learn common embeddings, the information from different layers in the sampled graph $\widetilde{G}$ is fused by applying the GAT \cite{c:13} method: 

\begin{equation}
    \label{eq:2}
    a_i={\rm softmax}(w_1 {\rm tanh}(W_2 H_i)),
\end{equation}

\begin{equation}
    \label{eq:3}
    H_i = (h_{i,1}, h_{i,2}, ..., h_{i,L}),
\end{equation}
where $w_1$, $W_2$ are a trainable vector and a trainable matrix, respectively, $H_i$ is a neighbor embedding set of $v_i$ and each $h_{i,l}$ is a neighbor embedding aggregated from the neighbors of $v_i$ on the $l$th layer. The $k$th-tier neighbor embedding $h_{i,l}^{(k)}$ is defined as:

\begin{equation}
    \label{eq:4}
    h_{i,l}^{(k)}  = {\rm AGG}_{mean}(\lbrace h_{j,l}^{(k-1)}, \forall v_j \in N_{i,l}, l=1,2,...,L \rbrace),
\end{equation}
where ${\rm AGG}_{mean}$ is a mean aggregator and $N_{i,l}$ is the neighbor set of $v_i$ on the layer $G_l$. For attributed multiplex networks, the initial neighbor embedding $h_{i,l}^{(0)}$ for $v_i$ in $G_l$ is defined as a parameterized function of $v_i$’s attributes \cite{c:5}, i.e., $h_{i,l}^{(0)}=f_l(r_i)$, where $f_l$ is a transformation function that transforms the features to a neighbor embedding. 

Thus, the common embedding $\widetilde{c}_i$ of $v_i$ is:

\begin{equation}
    \label{eq:5}
    \widetilde{c}_i = W_3H_ia_i,
\end{equation}
where $W_3$ is a trainable parameter matrix and $a_i$ is self-attention coefficients.

Finally, the overall embedding $\widetilde{z}_{i,l}$ of $v_i$ on the layer $\widetilde{G}_l$ is:

\begin{equation}
    \label{eq:6}
    \widetilde{z}_{i,l} = \widetilde{c}_i + \widetilde{x}_{i,l}.
\end{equation}

\subsection{Model Training}
For positive samples on the layer $\widetilde{G}_l$, the loss function is defined as:

\begin{equation}
    \label{eq:7}
    \mathfrak{L}_{pos}^l = \sum_{(v_i, v_j)\in \widetilde{E}_l} -{\rm log}(\sigma({\widetilde{z}_{i,l}}^T \widetilde{z}_{j,l})),
\end{equation}
where $\widetilde{E}_l$ is the edge set on the sampled layer $\widetilde{G}_l$, $\sigma$ is a nonlinear activation function and $\widetilde{z}_{i,l}$ and $\widetilde{z}_{i,l}$ are the final embeddings of $v_i$ and $v_j$ on the layer $\widetilde{G}_l$, respectively. The overall loss of positive samples is:

\begin{equation}
    \label{eq:8}
    \mathfrak{L}_{pos} = \sum_{l=1}^L \mathfrak{L}_{pos}^l.
\end{equation}

In order to better learn the node embeddings on the target layer, the negative samples are all selected from the target layer and the corresponding loss is defined as:

\begin{equation}
    \label{eq:9}
    \mathfrak{L}_{neg} = \sum_{(v_i, v_j)\in D_t} -{\rm log}(1-\sigma(\widetilde{z}_{i,t}^T \widetilde{z}_{j,t})),
\end{equation}\textbf{\textbf{}}
where $D_t$ is the negative sample set from the target layer $G_t$, $\widetilde{z}_{i,t}$ and $\widetilde{z}_{j,t}$ are the final embeddings of $v_i$ and $v_j$ on the layer $G_t$, respectively. 

The overall loss $\mathfrak{L}_{total}$ of LIAMNE is:

\begin{equation}
    \label{eq:10}
    \mathfrak{L}_{total} = \mathfrak{L}_{pos} + \mathfrak{L}_{neg}.
\end{equation}

\section{Experiment}
In this section, we empirically evaluate the performance of LIAMNE and six comparative methods on six real-world datasets. 
Two tasks, i.e., link prediction task on the sparse layer and node classification are used to demonstrate the effectiveness and robustness of our method, respectively. 
Furthermore, the ablation analysis of the sampling module and the sensitivity analysis of the hyper-parameters are presented.

\begin{table*}[!t]
\centering
\begin{tabular}{ccccccc}
\toprule
\makecell{Dataset}  & \makecell{Layers} & \makecell{Nodes} & \makecell{Edges@D} & 
\makecell{Edges@S} & \makecell{Imbalance ratio}&\makecell{Density@S($\times 10^{-5}$)}\\
\midrule
\makecell{FFTWYT} & \makecell{3} & \makecell{6,407} & 
\makecell{42,327} & \makecell{614} & \makecell{4.23} & \makecell{1.49} \\
\makecell{Sacch-Pomb} & \makecell{7} & \makecell{4,092} & 
\makecell{34,192} & \makecell{240} & \makecell{4.95} & \makecell{1.43} \\
\makecell{Sacch-Cere} & \makecell{7} & \makecell{6,570} & 
\makecell{109,045} & \makecell{1,426} & \makecell{4.33} & \makecell{3.30} \\
\makecell{Rattus} & \makecell{3} & \makecell{2,640} & 
\makecell{3,014} & \makecell{122} & \makecell{1.54} & \makecell{1.75} \\
\makecell{IMDB} & \makecell{2} & \makecell{3,550} & 
\makecell{66,428} & \makecell{13,788} & \makecell{1.57} & \makecell{109} \\
\makecell{IMDB*} & \makecell{2} & \makecell{3,550} & 
\makecell{50,484} & \makecell{811} & \makecell{4.13} & \makecell{6.43} \\
\makecell{DBLP} & \makecell{2} & \makecell{7,907} & 
\makecell{144,783} & \makecell{90,145} & \makecell{0.47} & \makecell{144} \\
\makecell{DBLP*} & \makecell{2} & \makecell{7,907} & 
\makecell{109,428} & \makecell{2,039} & \makecell{3.98} & \makecell{3.26} \\
\bottomrule
\end{tabular}
\caption{Statistics of the datasets. Edges@D represents the number of edges on the densest layer and Edges@S represents the number of edges on the sparsest layer (i.e, the target layer). The definition of imbalance ratio can be found in Definition 2. The density of the target layer is defined as $\frac{|E_t|}{|V|\times (|V|-1)}$
where $|E_t|$ is the number of edges on the target layer and $|V|$ is the number of nodes.
}
\label{tab:1}
\end{table*}

\subsection{Dataset}
The datasets used in the experiment cover multiple fields, including social networks, biological networks, and publication networks, 
with varying degrees of layer imbalance. The statistics of all datasets are shown in Table \ref{tab:1}. 

FFTWYT\footnote[1]{\url{http://multilayer.it.uu.se/datasets.html}\label{web:1}} \cite{c:17} is a social network that contains public interactions among users of Friendfeed, Twitter and YouTube. Three different layers include commenting, liking and following interactions, respectively.

Sacch-Pomb\footnote[2]{\url{https://manliodedomenico.com/data.php}\label{web:2}} and Sacch-Cere\textsuperscript{\ref{web:2}} \cite{c:18} are biological multiplex networks on Saccharomyces pombe and Saccharomyces cerevisiae, respectively. Both have seven layers of interactions: direct interactions, physical associations, suppressive genetic interactions, synthetic genetic interactions, and additive genetic interactions.

Rattus\textsuperscript{\ref{web:2}} \cite{c:18} is a subset of BioGRID concerning protein interactions of Rattus Norvegicus. It has three layers of interactions: physical associations, direct interactions, colocalizations.

IMDB\footnote[3]{\url{https://www.imdb.com/}\label{web:3}} is a movie network that has two types of movie relations: movie-actor-movie and movie-director-movie. The attribute of each movie is a 1,007-dimensional bag-of-words representation of its plot.

DBLP\footnote[4]{\url{https://aminer.org/AMinerNetwork}\label{web:4}} \cite{c:19} is a publication network containing two types of paper relations: paper-paper and paper-author-paper. The attribute of each paper is a 2,000-dimensional bag-of-words representation of its abstract.

DBLP* and IMDB* are two artificially constructed multiplex networks that creates a sparser target layer by random sampling to increase the layer imbalance ratio of the original DBLP and IMDB. The purpose of constructing these two datasets is to compare the results of node classification on balanced and imbalanced multiplex networks.

\subsection{Competitors}
The comparison models include two single-layer network embedding models node2vec and LINE, and four multiplex network embedding models CrossMNA, GATNE, DGMI and HDMI. 

node2vec \cite{c:2} designs a biased random walk and explores diverse neighborhoods to learn richer representations.

LINE \cite{c:16} uses both the $1^{st}$- and $2^{nd}$-order of node proximity to learn node representations.

CrossMNA \cite{c:9} leverages the cross-network information to refine two types of node embedding vectors, i.e., inter-vector for network alignment and intra-vector.

GATNE \cite{c:5} performs skip-gram over the node sequences generated by random walk on each layer to learn a base embedding and different types of edge embeddings for each node.

DMGI \cite{c:21} extends DGI \cite{c:20} onto multiplex networks and uses consensus regularization to combine node embedding from different layers.

HDMI \cite{c:22} splits a given multiplex network into multiple attributed graphs. For each of them, three different objectives are proposed to maximize the mutual information between raw node features, node embeddings, and graph-level representations.

\subsection{Experiment Configuration}
The node embedding dimension of all models is set as 64. For node2vec and LINE, we combine a multiplex network into a single-layer network and remove duplicate edges. For CrossMNA, inter-vector is used for node classification and intra-vector is used for link prediction. For GATNE, we use the overall node embeddings from the target layer for link prediction and the mean of overall node embeddings from different layers for node classification. Since DMGI and HDMI are proposed for attributed networks, we randomly generate node attributes for attribute-free multiplex networks in the link prediction task.

In LIAMNE, the dimensions of $\widetilde{c}$ and $\widetilde{x}_l$ are both set as 64. 
We apply $\alpha$=0.2 and $\beta$=0.6 for under-sampling and start sampling in the $2^{nd}$ epoch. 
In link prediction, the overall node embeddings of the target layer are calculated by $\widetilde{c}+\widetilde{x}_l$, 
which are still 64-dimensional vectors. In node classification, the mean of overall node embeddings from different layers is utilized.

\begin{table*}[!t]
\centering
\begin{tabular}{ccccccccc}
\toprule
\makecell{Dataset}  & \makecell{FFTWYT} & \makecell{Sacch-Pomb} & \makecell{Sacch-Cere} & 
\makecell{Rattus} & \makecell{IMDB} & \makecell{IMDB*} & \makecell{DBLP}&\makecell{DBLP*}\\
\midrule
\makecell{node2vec} & \makecell{0.6749} & \makecell{0.5440} & \makecell{0.4797} & \makecell{0.4363} 
& \makecell{0.8592} & \makecell{0.6474} & \makecell{\underline{0.9696}} & \makecell{0.7752} \\
\makecell{LINE} & \makecell{0.7288} & \makecell{0.7327} & \makecell{0.7618} & \makecell{0.6672} 
& \makecell{0.9434} & \makecell{0.5974} & \makecell{0.9687} & \makecell{0.6444} \\
\midrule
\makecell{CrossMNA} & \makecell{0.8129} & \makecell{\underline{0.8461}} & \makecell{0.7624} & \makecell{\underline{0.8000}}
& \makecell{0.8001} & \makecell{0.5238} & \makecell{0.9243} & \makecell{0.6036} \\
\makecell{GATNE} & \makecell{\underline{0.8316}} & \makecell{0.8095} & \makecell{0.7866} & \makecell{0.6323} 
& \makecell{\textbf{0.9960}} & \makecell{0.6639} & \makecell{\textbf{0.9875}} & \makecell{0.7814} \\
\makecell{DGMI} & \makecell{0.6741} & \makecell{0.7701} & \makecell{0.7844} & \makecell{0.6565} 
& \makecell{0.8180} & \makecell{0.6930} & \makecell{0.9069} & \makecell{\underline{0.8859}} \\
\makecell{HDMI} & \makecell{0.7497} & \makecell{0.6968} & \makecell{\underline{0.7975}} & \makecell{0.6603} 
& \makecell{0.9558} & \makecell{\underline{0.6993}} & \makecell{0.8809} & \makecell{0.8735} \\
\midrule
\makecell{LIAMNE(Ours)} & \makecell{\textbf{0.8372}} & \makecell{\textbf{0.8850}} & \makecell{\textbf{0.8541}} & \makecell{\textbf{0.8169}} 
& \makecell{\underline{0.9861}} & \makecell{\textbf{0.7428}} & \makecell{0.9345} & \makecell{\textbf{0.9052}} \\
\bottomrule
\end{tabular}
\caption{Performance comparison of AUC on link prediction.}
\label{tab:2}
\end{table*}

\subsection{Link Prediction}
\subsubsection{Training Settings.}
We apply each model to link prediction on the target layer, using AUC values for performance evaluation. We randomly select edges from the target layer to form training, validation and test sets with a ratio of 8:1:1. All models are trained three times to get the average results.
\subsubsection{Results.}
The results of link prediction are reported in Table \ref{tab:2}. It demonstrates that: 
1) our model achieves significantly higher AUC values than baselines especially on layer-imbalanced multiplex networks, which verifies that our under-sampling method can reduce the noisy edges with low correlations to the target layer, making the auxiliary information more efficient;
2) on DBLP and IMDB datasets with relatively balanced layers, LIAMNE still has a competitive performance, but is lower than GATNE, especially in DBLP. It is probably because 
our sampling method is ineffective on layer balanced networks, while the ramdom walk strategy adopted by GATNE can better capture global structural information on a relatively dense target layer;
3) HDMI and DMGI perform generally lower than other MNE methods on FFTWYT, Sacch-Pomb, Sacch-Cere and Rattus, 
suggesting that they are originally designed for attributed networks and thus may achieve poor performance on attribute-free datasets;
4) although node2vec and LINE perform better on layer-balanced and dense networks, 
their performance deteriorates significantly on layer-imbalanced datasets, indicating the poor robustness of single-layer network embedding methods.

\begin{figure}[h]
	\centering
	\includegraphics[width=230pt]{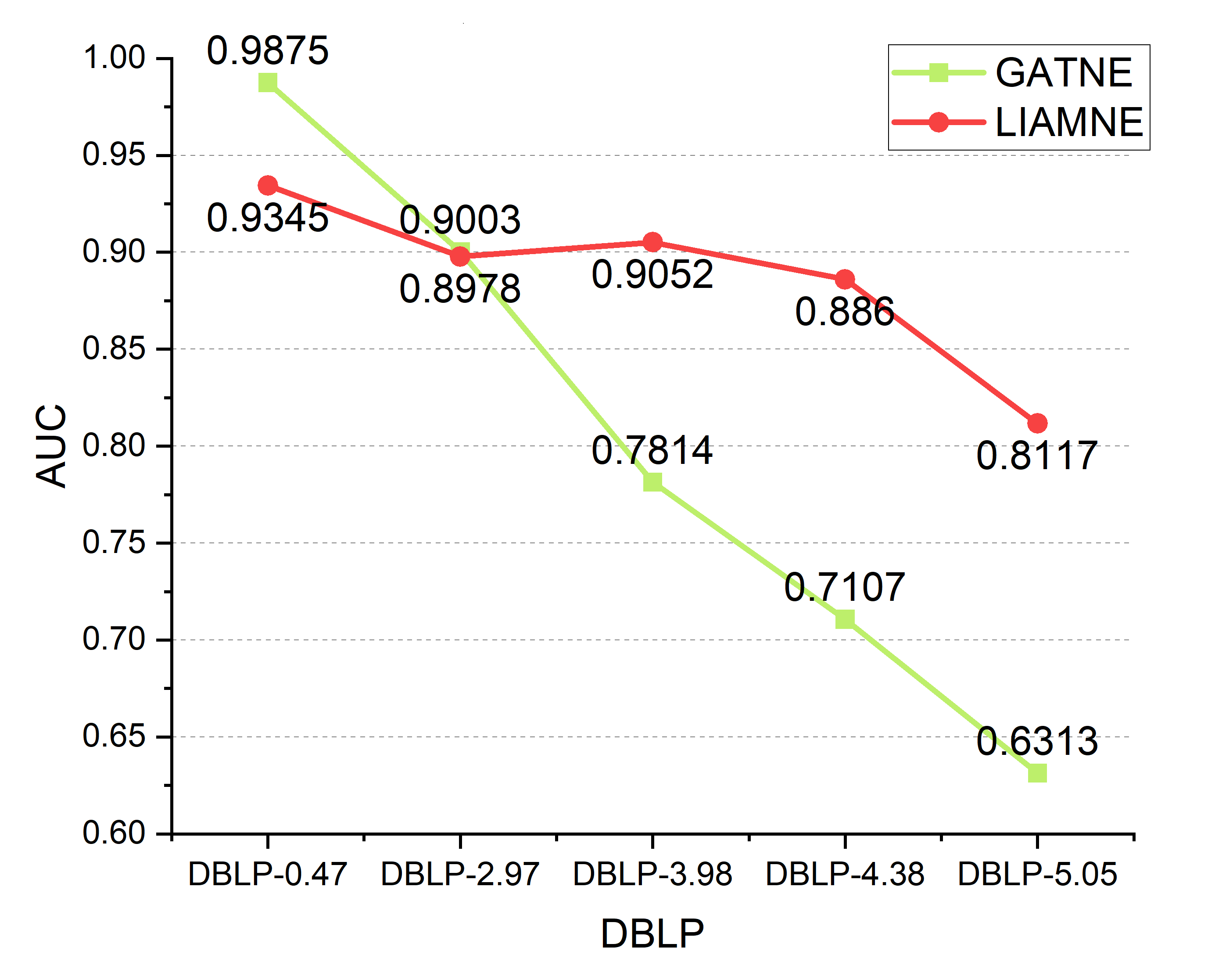}
	\caption{Performance of GATNE and LIAMNE with different imbalance ratios on DBLP}
	\label{fig:7}
\end{figure}

To further compare the performance of GATNE and LIAMNE on the same dataset with diverse layer imbalance ratios, we sparse the target layer of DBLP to varying degrees by random sampling to generate several new datasets and train the two models on them. Figure \ref{fig:7} shows the link prediction results of GATNE and LIAMNE on different DBLP datasets. It shows that the performance of GANTE drops sharply as the layer-imbalance ratio increases, indicating that the model may fail when the layers of the network are extremely imbalanced. As expected, the overall performance of LIAMNE is relatively stable, albeit slowly declining. 

\begin{table*}[!t]
\centering
\begin{tabular}{ccccccccc}
\toprule
\makecell{Dataset} & \multicolumn{2}{c}{IMDB} & \multicolumn{2}{c}{IMDB*} & \multicolumn{2}{c}{DBLP} & \multicolumn{2}{c}{DBLP*}\\
\midrule
\makecell{Metric} & \makecell{Macro F1} & \makecell{Micro F1} & \makecell{Macro F1} & \makecell{Micro F1} & 
\makecell{Macro F1} & \makecell{Micro F1} & \makecell{Macro F1} & \makecell{Micro F1} \\
\midrule
\makecell{node2vec} & \makecell{0.4854} & \makecell{0.4972} & \makecell{0.4749} & \makecell{0.4860} 
& \makecell{0.7444} & \makecell{0.7445} & \makecell{0.6884} & \makecell{0.6882} \\
\makecell{LINE} & \makecell{0.4968} & \makecell{0.5170} & \makecell{0.4442} & \makecell{0.4597} 
& \makecell{0.7207} & \makecell{0.7197} & \makecell{0.5861} & \makecell{0.5846} \\
\midrule
\makecell{CrossMNA} & \makecell{0.3457} & \makecell{0.3643} & \makecell{0.3295} & \makecell{0.3602} 
& \makecell{0.6912} & \makecell{0.7032} & \makecell{0.6267} & \makecell{0.6364} \\
\makecell{GATNE} & \makecell{0.6194} & \makecell{0.6219} & \makecell{0.5816} & \makecell{0.5921} 
& \makecell{0.8254} & \makecell{0.8177} & \makecell{0.7671} & \makecell{0.7508} \\
\makecell{DGMI} & \makecell{\underline{0.6505}} & \makecell{\textbf{0.6501}} & \makecell{\textbf{0.6546}} & \makecell{\textbf{0.6581}} 
& \makecell{0.7820} & \makecell{0.7775} & \makecell{0.8151} & \makecell{0.8013} \\
\makecell{HDMI} & \makecell{0.6408} & \makecell{0.6362} & \makecell{\underline{0.6391}} & \makecell{\underline{0.6432}} 
& \makecell{\underline{0.8318}} & \makecell{\underline{0.8244}} & \makecell{\underline{0.8249}} & \makecell{\underline{0.8114}} \\
\midrule
\makecell{LIAMNE(Ours)} & \makecell{\textbf{0.6509}} & \makecell{\underline{0.6453}} & \makecell{0.5531} & \makecell{0.5594} 
& \makecell{\textbf{0.8424}} & \makecell{\textbf{0.8378}} & \makecell{\textbf{0.8348}} & \makecell{\textbf{0.8302}} \\
\bottomrule
\end{tabular}
\caption{Performance comparison of Macro F1 and Micro F1 on node classification.}
\label{tab:3}
\end{table*}

\subsection{Node Classification}
\subsubsection{Training Settings.}
Then, we observe the representation ability across layers of our model for multiplex networks through the node classification task. We train a logistic regression classifier on the learned node embeddings. The ratio of the training set, validation set and test set is 8:1:1. We use Macro-F1 and Micro-F1 as the evaluation metrics of node classification.
\subsubsection{Results.}
Table \ref{tab:3} summarizes the node classification performance of all models on four datasets with node labels. 
The results show that: 1) overall, our model achieves competitive results on both layer-balanced and imbalanced datasets, 
indicating that our under-sampling method does not sacrifice the node representation of auxiliary layers when enhancing that of the target layer; 
2) as expected, attribute-aware multiplex network embedding methods, such as LIAMNE(ours), HDMI, DMGI and GATNE, 
generally perform better than those models that fail to leverage node attributes;
3) on IMDB* dataset, the performance of our method degrades. It can be observed that the node classification task on IMDB is inherently a difficult task and moreover the target layer of IMDB* is too sparse. As a result, it is more difficult for LIAMNE, which relies on the embedding accuracy of the target layer, to capture the information of auxiliary layers effectively, thus further deteriorating the comprehensive node embeddings of the whole network.

\begin{figure}[h]
\includegraphics[width=230pt]{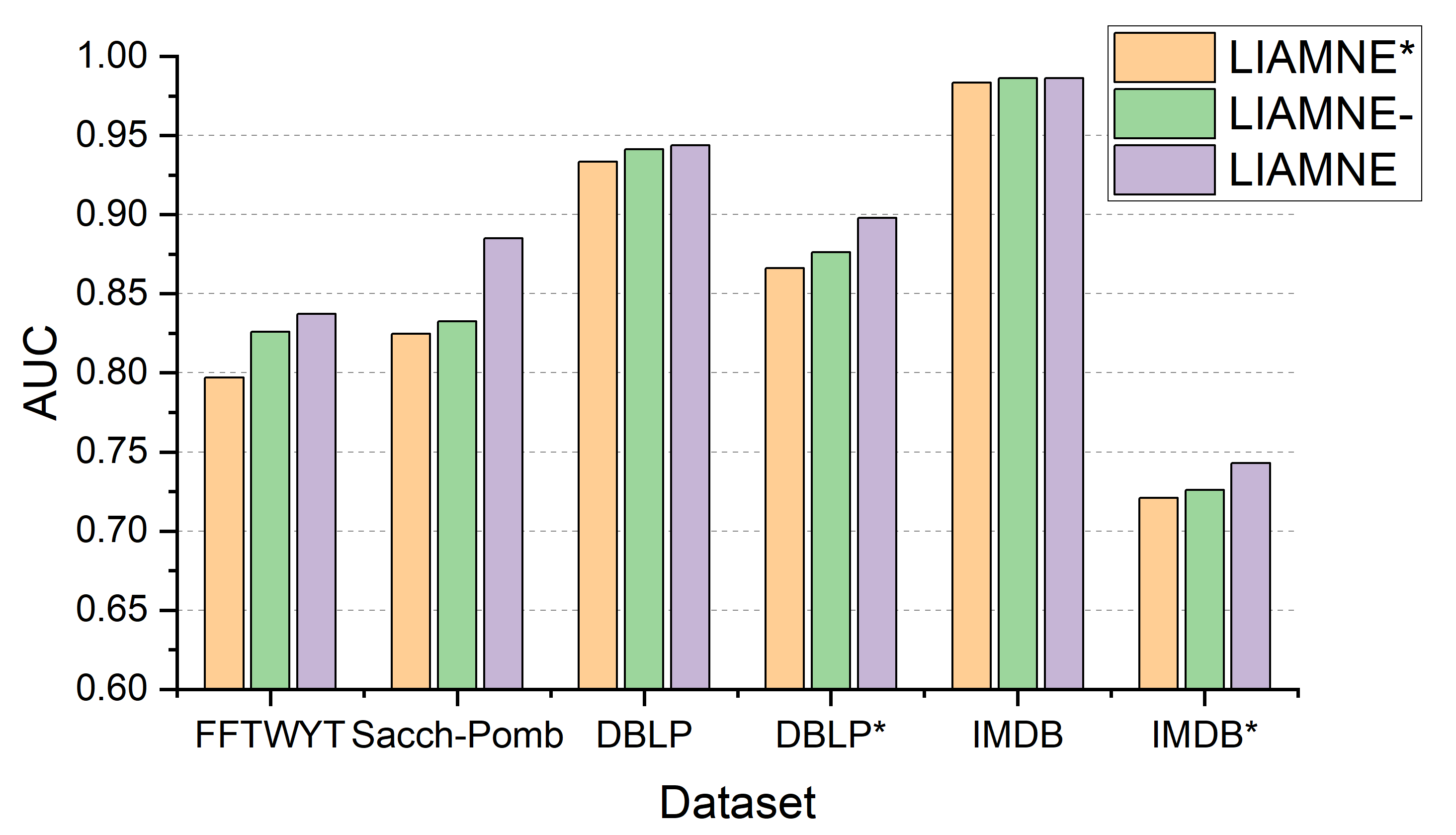}
\caption{Ablation study of LIAMNE*, LIAMNE- and LIAMNE.}
\label{fig:2}
\end{figure}

\subsection{Ablation analysis of under-sampling module}
To further measure the impact of the under-sampling module in LIAMNE, we conduct ablation studies using the following two model variants: LIAMNE-, which replaces the similarity-based sampling strategy with a random sampling strategy, and LIAMNE*, which completely removes the sampling module and leaves the rest unchanged. The comparative results in link prediction are shown in Figure \ref{fig:2}. It can be concluded that under-sampling does effectively improve the learning of node embedding on the sparse target layer and moreover the proposed under-sampling method based on the node similarity on the target layer is further verified to be significantly better than the random sampling method.
\\

\begin{figure}[h]
\includegraphics[width=230pt]{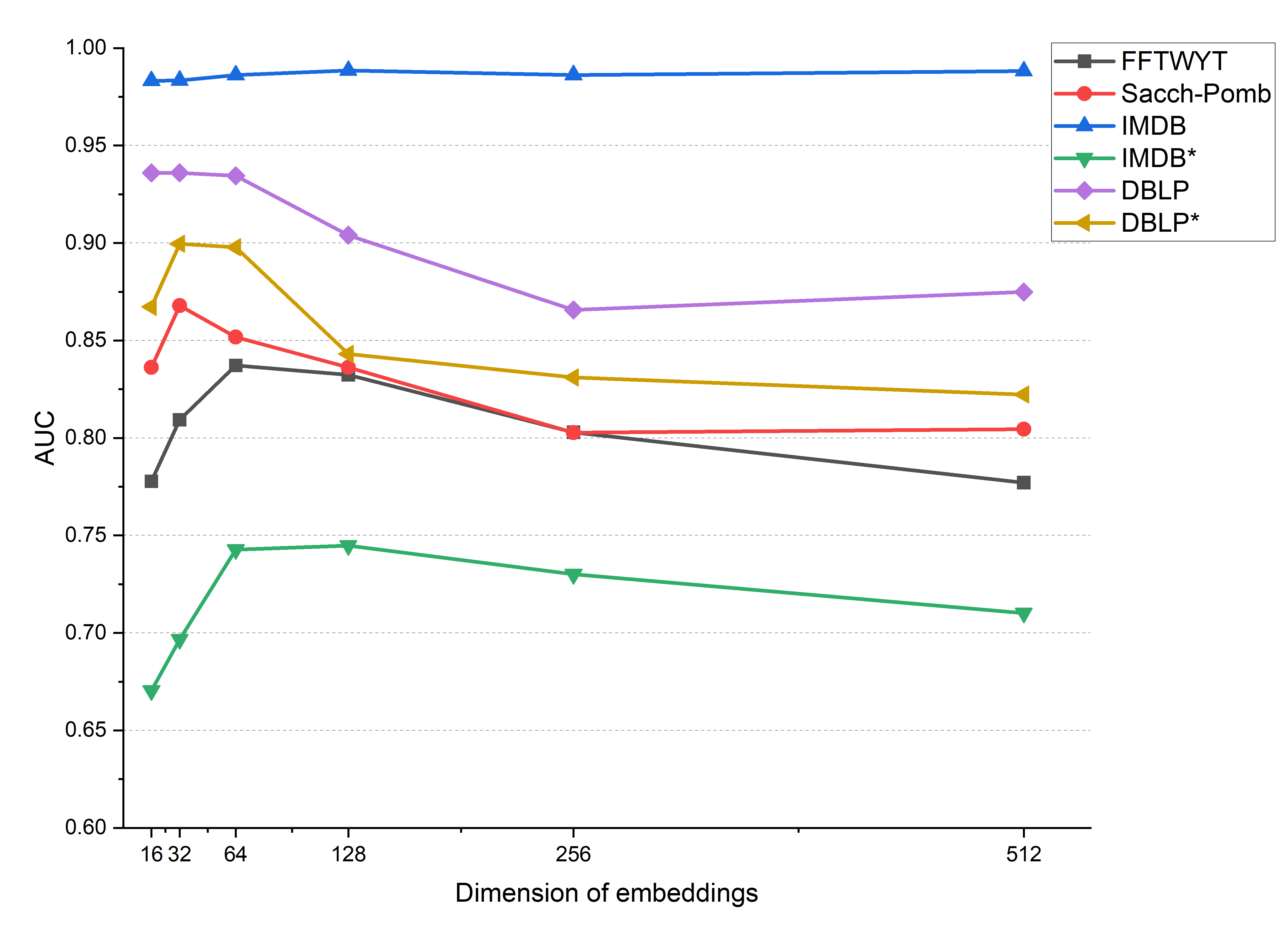}
\caption{Performance of LIAMNE when changing embedding dimensions.}
\label{fig:4}
\end{figure}

\begin{figure}[h]
\centering
\subfigure[AUC of FFTWYT]{\includegraphics[width=115pt]{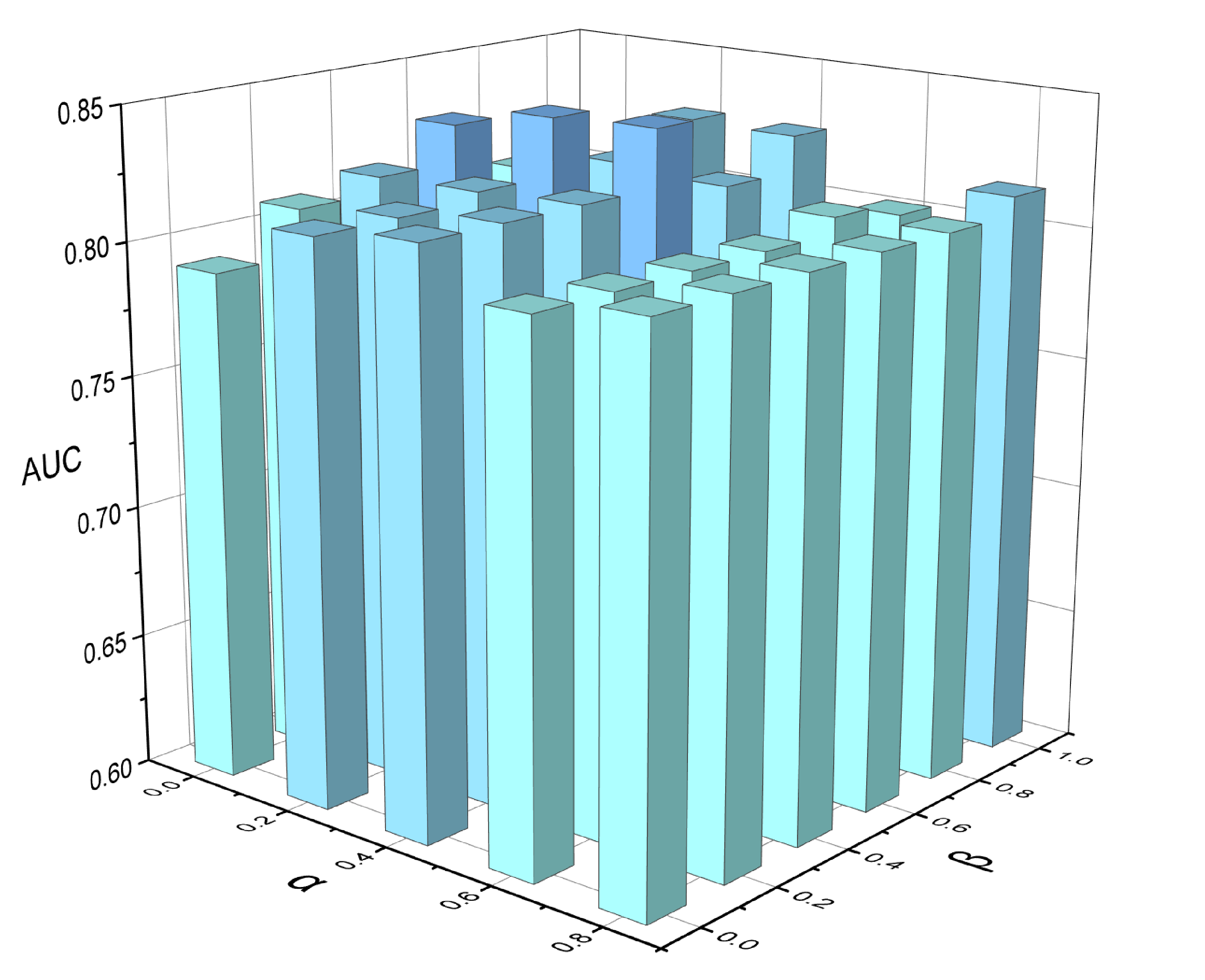}}
\subfigure[AUC of Sacch-Pomb]{\includegraphics[width=110pt]{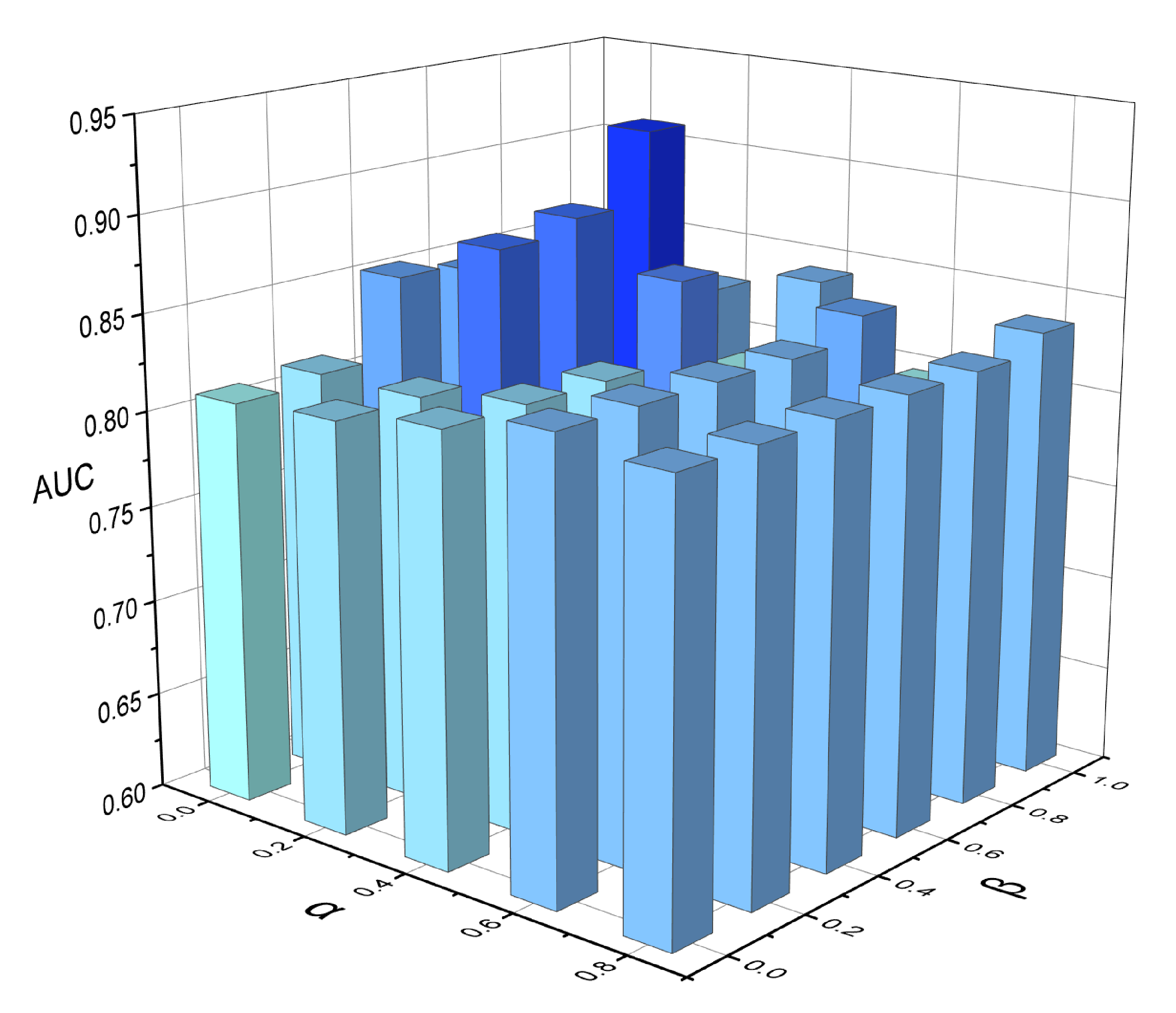}}
\caption{Performance comparison of LIAMNE with different $\alpha$ and $\beta$.}
\label{fig:5}
\end{figure}

\subsection{Analysis of Parameter Sensitivity}
In this section, we analyze three hyperparameters in LIAMNE, 
including embedding dimension $d$, similarity thresholds $\alpha$ and $\beta$. 
Figure \ref{fig:4} shows the results of link prediction under different embedding dimensions. 
We can conclude that the expected performance of LIAMNE will be obtained when $d\in[32, 128]$ and the performance drops when $d$ is either too small or too large. 
The optimal dimension may fluctuate slightly on different datasets, but the model always performs well when the dimension is 64, which becomes the default setting.

Figure \ref{fig:5} shows the AUC results of LIAMNE with different $\alpha$ and $\beta$ on FFTWYT and Sacch-Pomb. 
We observe that LIAMNE achieves ideal experimental results when $\alpha\in[0, 0.4]$ and $\beta\in[0.4, 0.8]$. 
This is because the important information from the auxiliary layer may be removed when $\alpha$ is too large. Meanwhile, the noise from auxiliary layers may not be effectively filtered out if the value of $\beta$ is set too small.
\\
\begin{figure}[!h]
	\includegraphics[width=230pt]{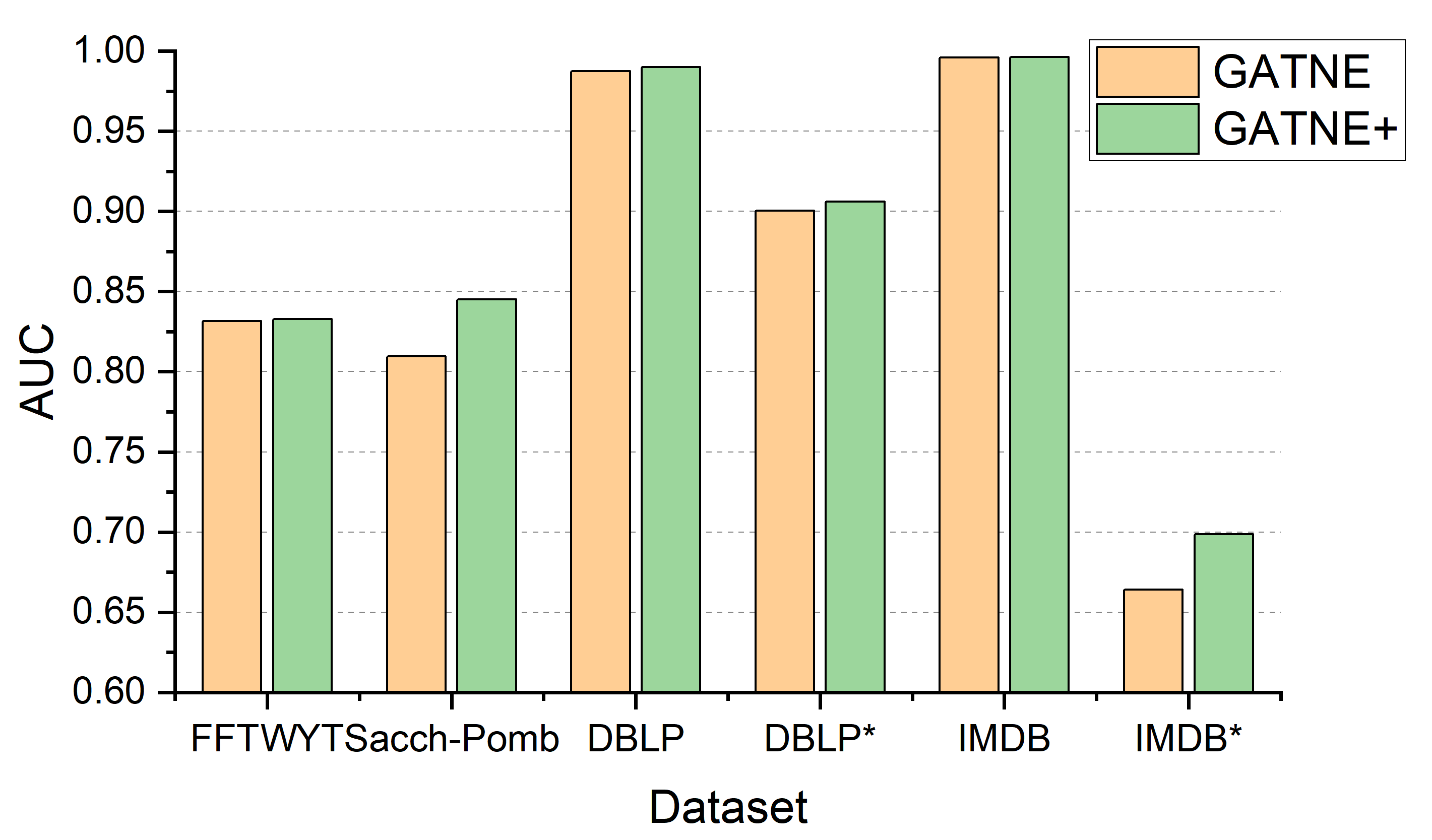}
	\caption{Performance comparison of GATNE and GATNE+.}
	\label{fig:3}
\end{figure}

\subsection{Portability of under-sampling module}
To test whether our under-sampling method is still effective to other MNE models, 
we apply it to GATNE to obtain GATNE+. Figure \ref{fig:3} illustrates the performance comparison of GATNE and GATNE+ in link prediction on the target layer. 
It can be seen that GATNE+ has significant improvement on Sacch-Pomb, DBLP* and IMDB* and has comparable results on FFTWYT, DBLP and IMDB. 
This suggests that our under-sampling method can be generally portable to enhance other MNE models.

\section{Conclusion}
This paper proposes an under-sampling method for multiplex network embedding, which effectively solves the problems of learning bias and data noise caused by layer imbalance. The method selectively samples the edges on auxiliary layers according to node distances in the embedding space of the target layer. The performance on link prediction on the sparse layer and node classification shows that the method can not only enhance the node embeddings from the sparse layer, but also make the overall embeddings of nodes more robust. For future work, more fine-grained layer imbalance metrics will be investigated and adaptive methods to address the layer imbalance problem will be explored.


\bibliography{aaai23}
\end{document}